\documentclass[12pt,showpacs,preprintnumbers,amsmath,aps,amssymb]{revtex4}

\usepackage{graphicx}
\usepackage{bm}
\begin{document}
\baselineskip 16pt

\title{Non Commutative  Bianchi type II Quantum Cosmology  }
\author{M. Aguero$^2$}
\email{magxim@gmail.com}
\author{J. A. Aguilar S.$^2$}
\email{polarequis@yahoo.com.mx}
\author{C. Ortiz$^{1}$}
\email{ortizca@fisica.ugto.mx}
\author{M. Sabido$^{1}$}
\email{msabido@fisica.ugto.mx}
\author{J. Socorro$^{1,2}$}
\email{socorro@fisica.ugto.mx}
\affiliation{$^1$ Instituto de F\'{\i}sica de la Universidad de Guanajuato,\\
 A.P. E-143, C.P. 37150, Le\'on, Guanajuato, M\'exico\\
 $^2$ Facultad de Ciencias de la Universidad Aut\'onoma del Estado de M\'exico,\\
Instituto Literario  No. 100, Toluca, C.P. 50000, Edo de Mex, M\'exico}%

\date{\today}

\begin{abstract}
 In this paper we present the noncommutative  Bianchi Class A cosmological models coupled to 
barotropic perfect fluid. The commutative and noncommutative 
quantum solution to the Wheeler-DeWitt equation for any  factor ordering, 
to the anisotropic Bianchi type II  cosmological model are found, using a stiff fluid ($\gamma=1$). In our toy model, 
we introduce noncommutative scale factors, is say, we consider that  all minisuperspace  variables $\rm q^i$ does not commute, 
so the simplectic structure was modified.
\end{abstract}

\pacs{04.60.Kz; 04.20.Fy, 04.20.Jb, 98.80.Hw, }
\maketitle

\section{ Introduction}
In the last few years there has been several attempts to study the possible 
effects of noncommutativity in the cosmological scenario.
In particular, Compean et al, (Compean et al,2002,2003a,b) avoid the difficult 
technicalities of analyzing noncommutative cosmological models, when these are 
derived from the full noncommutative theory of gravity, see by example
(Ba\~{n}ados et al, 2001; Nishino-Rajpoot,2002; Cacciatori et al, 2002,hep-th/0203038;
Abe-Nair,hep-th/0212270; Cardella-Zanon,hep-th/0212071; Chaichian et al, 2001),  
their proposal is to introduce the effects
of noncommutativity in quantum cosmology including one $\theta$ parameter,  deforming the minisuperspace. This 
deformation is achieved through a Moyal deformation of the
Wheeler-DeWitt (WDW) equation, this is similar to noncommutative quantum mechanics
(Gamboa, 2001; Chaichian et al, 2001).
Some work has been done in this direction, for example in (Barbosa-2005) the 
authors study the implications of noncommutative geometry in minisuperspace variables 
for a FRW universe with a conformally coupled scalar field, using the bohmian
formalism of quantum trajectories (Barbosa, 2004) also in (Pimentel, 2005) a
 noncommutative deformation of a scalar field coupled to
scalar-tensor type gravity was considered.

The aim of this paper is to build  a noncommutative scenario for the Bianchi
Class A cosmological models coupled with matter, we introduce noncommutative scale factors in Bianchi cosmologies
and compare their solutions to that of the commutative case at quantum level. Thus, the non commutativity is considered between all 
variables of  the minisuperspace (not in its momenta) with three $\theta_i$ parameters, $\rm \left[q^i, q^j \right]=i\theta^{ij}$, 
usually this is made by the Moyal star product $ \star$, however, we use one transformation between the non commutative
and the commutative  variables, resulting in a shifting, and it is well known that this correspondence, 
implies changes in the potential term of 
the WDW equation with a shifting in the variables (Compean et al, 2002; Pimentel, 2005). These transformation are not
the most general possible to define a noncommutative fields, see for example Carmona et al (Carmona, 2003), where the authors
give the generalization the noncommutative harmonic oscillator constructions and shown the most linear transformation between
the coordinates and momenta. In Vakili et al (Vakili, 2007) one procedure similar is applied to bianchi I without matter.
 
On the other hand, the inclusion of matter to  homogeneous cosmologies has been treated with scalar fields in order to study different
scenarios, inflation, dark matter, dark energy. However, since 1972 year,  
the problem of the appropriate sources of matter  pointed in (Ryan, 1972; Ryan-Shepley, 1975),
 and their corresponding lagrangian for each of them has not been solved. One of the 
 sources that is  usually considered,  is the  perfect fluid, in this paper we consider this source
as a first approximation in  the noncommutative quantization program, in 
particular for the Bianchi type II cosmological model is considered, as a toy model, with 
the idea to apply this procedure in a future to all Bianchi Class A models.

The paper is then organized as follows. In section II, we obtain  the WDW 
equation including the
barotropic matter  contribution.  In section III we present  the commutative 
quantum solutions for the cosmological  Bianchi type II,
and stiff fluid as the matter source,  section IV is devoted to the  non commutative 
quantum model and the noncommutative quantum solutions are presented. Final remarks 
are presented in Sec. V.

\section{The Wheeler-DeWitt equation}

Let us start by recalling  the canonical formulation of the ADM formalism for the
diagonal Bianchi Class A models. The metrics have the form
\begin{equation}
\rm ds^2= -(N^2- N^j N_j)dt^2 + e^{2\Omega(t)} e^{2\beta_{ij}(t)} \, \omega^i \omega^j,
\label{metrica}
\end{equation}
where $\rm N$ and $\rm N_i$ are the lapse and shift functions, respectively, $\Omega(t)$ is a scalar and
$\rm \beta_{ij}(t)$ a 3x3 diagonal matrix,
$\rm \beta_{ij}=diag(\beta_+ +\sqrt{3} \beta_-,\beta_+ -\sqrt{3} \beta_-, -2\beta_+)$,
$\rm \omega^i$ are one-forms that  characterize  each cosmological Bianchi
type model, and that obey $\rm d\omega^i= \frac{1}{2} C^i_{jk} \omega^j \wedge \omega^k,$
$\rm C^i_{jk}$ the structure constants of the corresponding invariance group 
(Ryan-Shepley, 1975). 

For the Bianchi type II, has the form
\begin{eqnarray}
\rm ds^2  &=& \rm - N^2dt^2 + e^{2\Omega}e^{-4\beta_+}dx^2 +
e^{2\Omega}e^{2\beta_+ +2\sqrt{3}\beta_-} dy^2 
+ e^{2\Omega}\left[x^2 e^{2\beta_+ +2\sqrt{3}\beta_-}
+ e^{2\beta_+ -2\sqrt{3}\beta_-} \right]dz^2 \nonumber\\
&&-\rm xe^{2\Omega}e^{2\beta_+ +2\sqrt{3}\beta_-}dydz -
xe^{2\Omega}e^{2\beta_+ +2\sqrt{3}\beta_-}dzdy.  \label{bi2}
\end{eqnarray}
The Lagrangian  is given by
\begin{equation}
 \rm L=\rm 6 e^{3\Omega} \left[\frac{\dot N \dot \Omega}{N^2} 
 -\frac{\ddot \Omega}{N} -2\frac{\dot \Omega^2}{N}
 -\frac{\dot \beta_+^2}{N} -\frac{\dot \beta_-^2}{N} 
  +\frac{1}{12}N e^{-2\Omega+4\beta_+ +4\sqrt{3}\beta_-}
  + \frac{8}{3} \pi G N \rho \right].
 \end{equation}
where we used a perfect fluid as the matter content in a comoving frame, see 
(Ryan, 1972; Pazos, 2000).
 When we rewrite this equation in a canonical way and using the solution to
  $\rm T^{\mu\nu}_{~~;\nu}=0$, we arrive to the Hamiltonian function
 \begin{equation}
 \rm H= \frac{e^{-3\Omega}}{24} \left[-P_\Omega^2 + P_+^2 + P_-^2 
 +12 e^{4(\Omega+\beta_++\sqrt{3}\beta_-)} 
 +384\pi G M_\gamma e^{-3(\gamma-1)\Omega} \right]=0,\nonumber
\label{hami1}
   \end{equation}
 and we assumed a barotropic state equation $\rm p=\gamma \rho$, where the parameter,
$\rm -1\leq \gamma \leq 1$.

By replacing $\rm P_{q^\mu}$
by $\rm -i \partial_{q^\mu}$ in (\ref {hami1}),
with $\rm q^\mu=(\Omega, \beta_+,\beta_-)$, we get the WDW equation.  
Following Hartle-Hawking (Hartle-Hawking, 1983) we introduce a semi-general factor 
ordering, this  gives
 \begin{equation}
 \rm \left[\frac{\partial^2}{\partial \Omega^2} - \frac{\partial^2}{\partial \beta_+^2} -
  \frac{\partial^2}{\partial \beta_-^2} +q\frac{\partial}{\partial \Omega}  
   +12 e^{4(\Omega+\beta_++\sqrt{3}\beta_-)} + 
   384\pi G M_\gamma e^{-3(\gamma-1)\Omega} \right]\Psi=0,\label{wdw1}
\end{equation}
that is equivalent to
\begin{equation}
   \rm \Box \Psi -q \frac{\partial \Psi}{\partial \Omega}- U(\Omega,\gamma)=0,\label{wdw2}
      \end{equation}
where $\rm \Box$ is the d'Alambertian in three dimensions with signature (-,+,+), and the 
potential function
$\rm U(\Omega,\gamma) =12 e^{4(\Omega+\beta_++\sqrt{3}\beta_-)} 
+384 \pi G M_\gamma e^{-3(\gamma-1) \Omega}$.

\section{Commutative Cosmological quantum solutions, for $\rm \gamma=1$  }
Now we consider a stiff fluid $\gamma=1$ as the matter content for our model, 
for the anisotropic commutative model. We simplify the equation (\ref{wdw2}), 
introducing the particular transformation  between the coordinates
\begin{equation}
\rm \xi=\Omega+\beta_+ +\sqrt{3}\beta_- , \quad \kappa=\Omega+\frac{\sqrt{3}}{3}\beta_-,
\quad \lambda=\Omega-2\beta_++\sqrt{3}\beta_- , \nonumber
\end{equation}
 having
\begin{equation}
\label{hamilvar}
\rm  q\frac{\partial\Psi}{\partial\xi}-3\frac{\partial^2\Psi}{\partial\xi^2}
+\left(12e^{4\xi} +384\pi G M_0\right)\Psi 
+\frac{2}{3}  \frac{\partial^2\Psi}{\partial\kappa^2}
+q\frac{\partial\Psi}{\partial\kappa}-6\frac{\partial^2\Psi}{\partial\lambda^2}+
q\frac{\partial\Psi}{\partial\lambda}=0.
\end{equation}
Using the separation variables method with
$\rm \Psi(\xi,\kappa,\lambda)=X(\xi)Y(\kappa)Z(\lambda)$ and  
sustituting in (\ref{hamilvar}), we obtain the set of differential equations
\begin{eqnarray}
\label{sisb2}
&&\rm-\frac{3}{X}\frac{d^2X}{d\xi^2}+\frac{q}{X}\frac{dX}{d\xi}+\left(12e^{4\xi} 
+C_0\right)=-12\mu^2, \label{x} \\
&&\rm\frac{2}{3Y}\frac{d^2 Y}{d\kappa^2}+\frac{q}{Y}\frac{dY}{d\kappa}=-\frac{2}{3}b_1^2, 
\label{y}\\
&&\rm-\frac{6}{Z}\frac{d^2 Z}{d\lambda^2}+\frac{q}{Z}\frac{dY}{d\lambda}=6c_1^2, \label{z}
\end{eqnarray}
with $\rm C_0=384\pi G M_1$, and for simplicity we choose the constants 
$\rm \mu, b_1, c_1$ in such away that satisfying the relation
between them as $\rm \frac{2}{3}b_1^2-6c_1^2=-12\mu^2$. 

Introducing the change of variables $\rm  z= e^{4\xi} \label{zeta}$ in equation 
(\ref{x}), we get an ordinary Bessel differential equation for z
\begin{equation}
\rm z^2 \frac{d^2 X}{dz^2} -\left(1+\frac{q}{12}  \right)z \frac{dX}{dz}-
\left(\frac{z}{4} +b_0 \right) X=0,
\end{equation}
where $\rm b_0=\frac{1}{48}(C_0+12\mu^2)$, so the physical  solution is
\begin{equation}
\rm X(\xi)_{\nu}=e^{q\xi} \, K_{\nu}\left( e^{2\xi} \right), 
\end{equation}
with $\rm \nu= \frac{1}{2}\sqrt{4\mu^2+128\pi G M_1 +q^2}$.

The solutions for the other two equations are
\begin{eqnarray}
\label{sisgrie}
\rm Y(\kappa)&=&\rm e^{-\frac{3q}{4}\kappa}\left[A_1e^{ip_1\kappa} +A_2e^{-ip_1\kappa}\right],\\
\rm Z(\lambda)&=&\rm e^{\frac{q}{12}\lambda}\left[B_1e^{ip_2\lambda} +B_2e^{-ip_2\lambda}\right], 
\end{eqnarray}
where $\rm p_1=\sqrt{b_1^2-\frac{9q^2}{16}}$ and $\rm p_2=\sqrt{c_1^2-\frac{q^2}{144}}$.

Finally, we find the wave function
\begin{equation}
\rm \Psi_{C_\nu}= e^{4\xi + q(\xi -\frac{3}{4} \kappa +\frac{1}{12}\lambda)}\, K_{\nu}\left(e^{2\xi}   \right)
 \left[A_1e^{ip_1 \kappa} +A_2e^{-ip_1 \kappa}\right] 
 \rm \left[B_1e^{ip_2 \lambda} +B_2e^{-ip_2 \lambda}\right]. 
\end{equation}
For the  factor ordering $q=0$, the solution is simplified.\\
\section{Non Commutative Cosmological quantum solutions }

Finally  we can proceed to the  non commutative model, actually  we will consider, 
that the minisuperspace  variables $\rm q^i$ do not commute, so the simplectic 
structure is modified as follows
\begin{equation}
\rm [q^i, q^j ]= i\theta^{ij}, \qquad [P_i, P_j]=0, \qquad [q^i,P_j]=i\delta^i_j, \qquad q^i=(\Omega,\beta_+,\beta_-),
\label{rules}
\end{equation}
in particular, we choose the following representation
\begin{equation}
\rm [\Omega, \beta_+ ]= i\theta_1, \quad [\Omega,\beta_- ]= i\theta_2, \quad [\beta_-,\beta_+ ]= i\theta_3
\label{rules1}
\end{equation}
where the $\rm \theta_i$ parameters are a measure of the non commutativity between the minisuperspace variables. 
The commutation relation (\ref{rules}) or (\ref{rules1}) are not the most general ones to define a noncommutative field.

It is well known, that this
non-commutativity can be formulated in term of non-commutative
minisuperspace functions with the Moyal star product $\star$ of
functions
\begin{eqnarray}
\rm f(\Omega,\beta_+) \star g(\Omega,\beta_+)&=&f(\Omega,\beta_+) \,
e^{i\frac{\theta_1}{2}(\overleftarrow{\partial}_\Omega
\overrightarrow{\partial}_{\beta_+}
-\overleftarrow{\partial}_{\beta_+}\,\overrightarrow{\partial}_\Omega
)} g(\Omega,\beta_+), \nonumber\\
\rm f(\Omega,\beta_-) \star g(\Omega,\beta_-)&=&f(\Omega,\beta_-) \,
e^{i\frac{\theta_2}{2}(\overleftarrow{\partial}_\Omega
\overrightarrow{\partial}_{\beta_-}
-\overleftarrow{\partial}_{\beta_-}\,\overrightarrow{\partial}_\Omega
)} g(\Omega,\beta_-), \\
\rm f(\beta_-,\beta_+) \star g(\beta_-,\beta_+)&=&f(\beta_-,\beta_+) \,
e^{i\frac{\theta_3}{2}(\overleftarrow{\partial}_{\beta_-}
\overrightarrow{\partial}_{\beta_+}
-\overleftarrow{\partial}_{\beta_+}\,\overrightarrow{\partial}_{\beta_-}
)} g(\beta_-,\beta_+)\nonumber
\end{eqnarray}

On the other hand, it is well known that this correspondence, implies changes in the potential term of 
the WDW equation with a shift in the variables (Compean et al, 2002; Pimentel, 2005), and one possibility  for recover
(\ref{rules1}), and mantaining the old relations to the noncommutative fields, it become
\begin{equation}
\rm \Omega_{nc} \to \Omega + \frac{\theta_1}{2}P_+ +\frac{\theta_2}{2} P_-, \quad \beta_{-nc}\to \beta_- -\frac{\theta_2}{2} P_\Omega+\frac{\theta_3}{2}P_+,
\quad \beta_{+nc} \to \beta_+-\frac{\theta_1}{2} P_\Omega - \frac{\theta_3}{2}P_-,
\label{transformation}
\end{equation}
but, these shifting modified the potential term as
\begin{equation}
\rm U(\Omega,\beta_\pm,\theta_i) =12 e^{4\left[\Omega +\beta_+ +\sqrt{3}\beta_- 
+\frac{i\theta_1}{2}(\frac{\partial}{\partial \Omega}-\frac{\partial}{\partial \beta_+})
+ \frac{i\theta_2}{2}(\sqrt{3}\frac{\partial}{\partial \Omega}-\frac{\partial}{\partial \beta_-})-
\frac{i\theta_3}{2}(\sqrt{3}\frac{\partial}{\partial \beta_+}-\frac{\partial}{\partial \beta_-})\right] } +C_0.
\end{equation}

Now we can construct the noncommutative WDW equation (NCWDW)  
\begin{eqnarray}
 \rm   &&\left[\frac{\partial^2}{\partial \Omega^2} - \frac{\partial^2}{\partial \beta_+^2}
  -  \frac{\partial^2}{\partial \beta_-^2} +q\frac{\partial}{\partial \Omega} + C_0 \right.  \nonumber\\
  && \rm  \left. + 12 e^{4\left[\Omega +\beta_+ +\sqrt{3}\beta_- 
  +\frac{i\theta_1}{2}(\frac{\partial}{\partial \Omega}-\frac{\partial}{\partial \beta_+})
+ \frac{i\theta_2}{2}(\sqrt{3}\frac{\partial}{\partial \Omega}-\frac{\partial}{\partial \beta_-})
-\frac{i\theta_3}{2}(\sqrt{3}\frac{\partial}{\partial \beta_+}-\frac{\partial}{\partial \beta_-})\right] } \right]\Psi=0.
\label{wdw3}
\end{eqnarray}
Using the generalized Baker-Campbell-Hausdorff formula (Wilcox, 1967; Garc\'{\i}a, 2006)
\begin{equation}
\rm e^{\eta(\hat A + \hat B)}= e^{-\eta^2[\hat A , \hat B]}\, e^{\eta \hat A} \, 
e^{\eta  \hat B}
\end{equation}
and the relation between the variables (\ref{rules1}), we obtain  in the new variables 
$\rm (\xi,\kappa,\lambda)$ 
\begin{eqnarray}
\label{newvar}
&&\rm  q\frac{\partial\Psi}{\partial\xi}-3\frac{\partial^2\Psi}{\partial\xi^2}+
\left(12e^{4\xi} e^{2i\theta_1 \frac{\partial}{\partial \kappa}} e^{6i\theta_1 \frac{\partial}{\partial \lambda}}
e^{\frac{4\sqrt{3}}{3}i\theta_2 \frac{\partial}{\partial \kappa}} 
e^{\frac{2\sqrt{3}}{3}i\theta_3 \frac{\partial}{\partial \kappa}} e^{6\sqrt{3}i\theta_3 \frac{\partial}{\partial \lambda}}
 +C_0\right) \Psi \nonumber\\
&& \rm +\frac{2}{3}  \frac{\partial^2\Psi}{\partial\kappa^2}
+q\frac{\partial\Psi}{\partial\kappa}-6\frac{\partial^2\Psi}{\partial\lambda^2}+
q\frac{\partial\Psi}{\partial\lambda}=0.
\end{eqnarray}

Now we can look for solutions, using the particular anzats 
$\rm \Psi= X(\xi) e^{(-\frac{3q}{4} \pm i p_1)\kappa} e^{(\frac{q}{12}\pm i p_2) \lambda}$,
on equation (\ref{newvar}), and taking in account that 
$\rm e^{i\theta \frac{\partial}{\partial x}} e^{\eta x}\equiv e^{i\eta \theta}e^{\eta x}$
yields the equation for the function $\rm X(\xi)$
\begin{equation}
  \rm -3\frac{d^2 X}{d \xi^2}+ q\frac{d X}{d\xi} 
+  \left(12e^{f(\theta_i)}e^{4\xi} 
 +C_0+12 \mu^2 \right)X=0, 
 \label{xx}
 \end{equation}
where the function  $\rm f(\theta_i)$ have the following structure
\begin{equation}
\rm f(\theta_i)=\theta_1\left[-i q \mp ( 2p_1+ 6 p_2)\right] + \theta_2\left(-i \sqrt{3} q \mp  \frac{4\sqrt{3}}{3}p_1\right)
 \mp \theta_3 \sqrt{3}\left( \frac{2}{3}p_1+ 6 p_2 \right), 
 \label{ftheta}
 \end{equation}
  and  the constant $\rm b_1, c_1$ and $\rm \mu$ mantain the same relation as in the 
commutative case. We can see that the equations
 (\ref{x}) and (\ref{xx}) have the same structure, then the corresponding solutions 
 to (\ref{xx}) become the modified Bessel function
 \begin{equation}
 \rm {X_{NC}}_\nu (z)= e^{q\xi} K_{\nu}\left(e^{ f(\theta_i)} e^{2\xi}\right),
 \end{equation}
 
Finally, we arrive to the noncommutative wave function
\begin{eqnarray}
\rm \Psi_{{NC}_\nu}=\rm e^{4\xi + q(\xi -\frac{3}{4} \kappa +
\frac{1}{12}\lambda)}\, K_{\nu}\left(e^{ f(\theta)} e^{2\xi}   \right)
e^{\pm ip_1 \kappa}e^{\pm ip_2 \lambda}.
\label{solutionnc}
\end{eqnarray}
  We can remark that the non commutative
     correction the Bessel function gives the commutative solutions when the parameters $\theta_i \to 0$ or for particular
     choose between the parameters in such a way that $\rm f(\theta_i)=0$.  

\section{Conclusions}

In this work 
we introduce noncommutative scale factors, is say, consider that  all minisuperspace  variables $\rm q^i$ does not commute, 
so the simplectic structure was modified. 
We have extended the non commutativity between the minisuperspace variables (fields) to three parameters as
\begin{equation}
\rm [\Omega, \beta_+ ]= i\theta_1, \quad [\Omega,\beta_- ]= i\theta_2, \quad [\beta_-,\beta_+ ]= i\theta_3,
\end{equation}
Also we present one particular transformation, eq. (\ref{transformation}), where these relations are hold, but does not unique, and are not
the most general ones to define a noncommutative fields.
Also we have found  the commutative and noncommutative quantum solutions 
for the Bianchi type II cosmological model, considering 
a  scenario dominated by stiff fluid. The noncommutative quantum solution have differents structure with respect to
the commutative solutions, modified by the function $\rm f(\theta_i)$, eq. (\ref{ftheta}). The 
 commutative solutions is recover if the parameters $\theta_i \to 0$ or for particular
     choose between the parameters in such a way that $\rm f(\theta_i)=0$.  This procedure will be
     applied to other Bianchi Class A models, and the results will be reported elsewhere.

\acknowledgments{This work was supported in part by CONACyT grant 47641 and Promep grant 
UGTO-CA-3.
Many calculations where supported by Symbolic Program REDUCE 3.8. }

\end{document}